\documentclass[aps,twocolumn,showpacs]{revtex4}
\usepackage{dcolumn}
\usepackage{graphicx}
\usepackage{amsmath}
\usepackage{amsfonts}
\usepackage{amssymb}
\usepackage{psfrag}
\usepackage{wrapfig}
\usepackage{subfigure}
\usepackage{makeidx}
\usepackage{bm}
\usepackage{epsf}

\begin{document}

\title{W-shaped solitons generated from a weak modulation in the Sasa-Satsuma equation}

\author{Li-Chen Zhao$^{1}$}\email{zhaolichen3@163.com}
\author{Sheng-Chang Li$^{2}$}\email{scli@mail.xjtu.edu.cn}
\author{Liming Ling$^{3}$}\email{linglm@scut.edu.cn}

\address{$^1$Department of Physics, Northwest University, 710069, Xi'an, China}
\address{$^2$School of Science, Xi'an Jiaotong University, 710049, Xi'an, China}
\address{$^3$Department of Mathematics, South China University of Technology, 510640, Guangzhou, China}
%%%%%%%%%%%%%%%%%%%%%%%%%%%%%%%%%%%%%%%%%%%%%%%%%
\date{\today}
\begin{abstract}
We revisit on rational solution of Sasa-Satsuma equation, which can be used to describe evolution of optical field in a nonlinear fiber with some high-order effects. We find a striking dynamical process which involves both modulational instability and modulational stability regimes, in contrast to the rogue waves and W-shaped soliton reported before which involves modulational instability and stability respectively. It is demonstrated that stable W-shaped solitons can be generated from a weak modulation signal on continuous wave background. This provides a possible way to obtain stable high-intensity pulse from low-intensity continuous wave background.

\end{abstract}
 \maketitle

\section{Introduction}
 Recently, rational solutions of nonlinear Schr\"{o}dinger
equation (NLSE) have been paid much attention, since  rational solution of NLSE can be used to describe rogue wave (RW) dynamics in many different physical systems \cite{report, Solli,Kibler,Bailung, Chabchoub,Chabchoub2,Chabchoub3}. Among these different systems, optical fiber play an important role in experimental observations for its well-developed intensity and phase modulation techniques. The experimental studies in nonlinear fiber showed that the simplified NLSE can well describe the dynamics of localized waves, which only contains
the group velocity dispersion (GVD) and its counterpart, namely, self-phase modulation (SPM). But for ultrashort pulses whose duration is shorter than $100$fs, which is tempting and desirable to improve the capacity of high-bit-rate transmission systems, the nonlinear susceptibility will produce higher-order nonlinear
effects like the Kerr dispersion (i.e., self-steepening) and the delayed nonlinear response except for SPM, and even the third-order dispersion (TOD). These are the most general terms that have to be
taken into account when extending the applicability of the NLSE \cite{K. Porsezian,Kodama}. With these effects, the corresponding
integrable equation was derived as Sasa-Satsuma(S-S) equation \cite{SS}.

The linearized stability analysis for the S-S equation in Ref. \cite{Wright} suggested that there are both modulational instability(MI) and modulational stability(MS) regimes for low perturbation frequencies on the continuous wave  background(CWB). In the MI regime, the rational solutions are obtained in  \cite{rws,Chen}, which correspond to RW excitation. It is demonstrated that the high-order effects could make the RW twisted and the rational solution of the S-S equation had distinctive properties in contrast to that of the well-known NLSE. In the MS regime, we presented rational solutions in the MS regime, which correspond to W-shaped soliton excitation \cite{Zhao}. It is shown that the profile of W-shaped soliton depends on the background frequency. The results suggested that not all rational solution of nonlinear partial equation correspond to RW dynamics. It should be noted that there are two critical points for background frequency on boundary of MI and MS regime. We expect that there should be some new dynamics on the critical points.

In this paper, we revisit on rational solution of S-S model on the critical background frequencies. We find two W-shaped solitons can be generated from a weak modulation signal on continuous wave background. The striking process should involves both modulational instability and modulational stability regimes, in contrast to the rogue waves and W-shaped soliton reported before which involves modulational instability and stability respectively.

\section{Two W-shaped solitons generated from a weak modulation on CWB}

For ultrashort pulses whose duration is shorter than $100$fs,  the optical field [i.e., $E(t,z)$] in a nonlinear fiber with some high-order effects can be described by the S-S model \cite{SS}:
\begin{eqnarray}\label{S-S}
&&i E_z+\frac{1}{2}E_{tt}+|E|^2E\nonumber\\&&+i\epsilon[E_{ttt}+ 3(|E|^2 E)_{t}+3
|E|^2 E_t]=0,
\end{eqnarray}
where  $\epsilon$ is an arbitrary real parameter to scale the integrable perturbations of the NLSE and it can change the strength of the high-order effects conveniently. The last three high-order terms  correspond to the TOD, self-steepening, and delayed nonlinear response respectively  \cite{Agrawal,Li,Porsezian,Kodama}. Raman effect  $T_R E |E|^2_t $ ($T_R$ is a real constant) is not considered in the model, since the term can be much smaller than Kerr dispersion effect, and it makes the pulse energy non-conservation (this usually makes the model non-integrable).  When $\epsilon=0$, the S-S equation reduces to the standard NLSE which has only the terms describing the lowest order dispersion and self-phase
modulation. The investigations on the S-S equation indicated that the nonlinear waves in nonlinear fiber with the high-order effects are much more diverse than the
ones in simplified NLSE \cite{Li,Porsezian,Mihalache,Mihalache2,Mihalache3}. Here, we revisit on rational solution of the S-S model.
\begin{figure}[t]
\centering{\includegraphics[width=0.8\columnwidth]{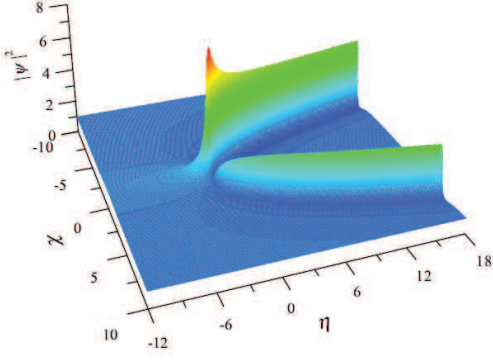}}
\caption{(Color online) The evolution of $|\psi|^2$ [given by Eq. (\ref{solution})] corresponding to the process that one weak modulation signal evolves into two W-shaped solitons. The parameters are $a=\gamma_1=\gamma_2=1$.}
\end{figure}

For convenience, let $\psi(T,Z)=E(t,z)\exp{[-\frac{i}{6 \epsilon}(t-\frac{z}{18\epsilon})]}$ \cite{Nakkeeran},
where $T=t-\frac{z}{12\epsilon}$ and $Z=z$, then a simplified equation equivalent with Eq. (\ref{S-S}) is obtained as $\psi_{Z}+\epsilon
(\psi_{TTT}+ 6 |\psi|^2 \psi_{T}+3 \psi|\psi|^2_T)=0$. The evolution direction of nonlinear waves in this equation is slant, which make us inconvenient to observe the dynamical behaviors. To overcome this difficulty, we introduce the coordinates transformation, namely, $\chi=a\left(T-\frac{33}{4}a^2 \epsilon Z\right)$ and $\eta=-\frac{3\sqrt{3}}{2}a^3\epsilon Z$, then the simplified equation can be written as
\begin{equation}\label{S-S-1}
    \psi_{\eta}=-\frac{11\sqrt{3}}{6}\psi_{\chi}+\frac{2\sqrt{3}}{9}\psi_{\chi\chi\chi}+\frac{2\sqrt{3}}{3a^2}(|\psi|^2\psi)_{\chi}
    +\frac{2\sqrt{3}}{3a^2}|\psi|^2\psi_{\chi}.
\end{equation}
The coordinates transformation only lead to a trivial phase multiplication from Eq. (\ref{S-S-1}) to Eq. (\ref{S-S}) and thus the identical dynamics are kept well. Subsequently, we will discuss the nonlinear dynamics of Eq. (\ref{S-S-1})  instead of Eq. (\ref{S-S}) without losing generality. On the CWB $\psi_0(\chi,\eta)=a \exp{[i \omega_0 \chi- i(\frac{\sqrt{3}}{2} \omega_0+\frac{2\sqrt{3}}{9} \omega_0^3) \eta]}$ with arbitrary $\omega_0$, it is convenient to study various localized waves such as RW with two peaks \cite{Chen,rws} and rational W-shaped soliton \cite{Zhao}. Surprisingly, in the case of $\omega_0=\pm \frac{1}{2}$, we find a dynamics process that one weak modulation signal on the background can evolve into two stable nonlinear waves as shown in Fig. 1. This process is very different from the dynamics of nonlinear waves for the same model reported in Refs. \cite{Chen, rws, Zhao}. From Fig. 1 we see that the propagating signal at the locations before $\eta=-2$ keeps small while it evolves to a large localized wave near $\eta=-1.5$. In particular, this large localized wave splits into two quasi-stabilized nonlinear waves after $\eta=-1.5$. The profiles of them can be observed clearly with the help of making cut plot at a certain location $\eta$, which show a nice W-shaped soliton configuration similar to the one reported in Ref. \cite{Zhao}. It is found that the higher peak of the two waves decreases with increasing the propagation distance $\eta$ while the lower one increases. However, after a certain distance, the changing rates of them will tend to zero, namely, the nonlinear localized waves tend to solitons whose shapes are kept well. This process could suggest one way to excite quasi-stabilized W-shaped soliton through adding small modulation on the continuous background, which has significant and potential values in the field of soliton application.

\begin{figure}[t]
\centering{\includegraphics[width=0.8\columnwidth]{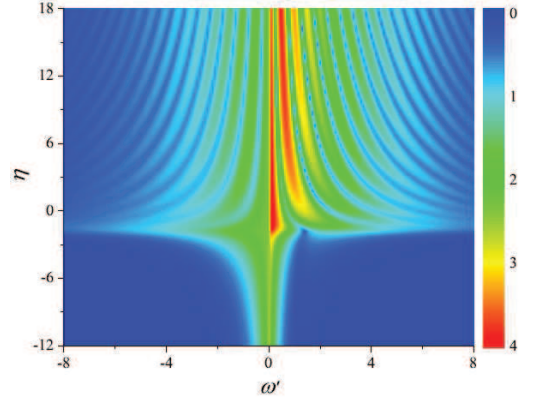}}
\caption{(Color online) The corresponding spectrum evolution of the process illustrated in Fig. 1. It is seen that there is an interference pattern on spectral distribution with $\omega'=\omega-\omega_0$. }
\end{figure}

The above dynamical process can be faithfully described by the following analytical solution
on the CWB with frequency $\omega_0=\pm\frac{1}{2}$, which can be obtained by adopting Darboux transformation method.
\begin{equation}\label{solution}
\psi(\chi,\eta)=a \left[1+\frac{H}{M}\right] \exp{\left[i(\pm \frac{\chi}{2}\mp \frac{5\sqrt{3}}{18}\eta)\right]},
\end{equation}
with
\begin{eqnarray}
H&=&H_1(\chi,\eta)+(2i\sqrt{3}-12i\chi-12i\gamma_2-6)\gamma_1\nonumber\\&&+
 [ 18i\chi^2-(18i\sqrt{3}+18)\chi-12i\eta+6\sqrt{3}+6i ]\gamma_2\nonumber\\&&+(12i\chi-8i\sqrt{3}-12)\gamma_2^2,\nonumber
 \end{eqnarray}
\begin{eqnarray}
 M&=&M_1(\chi,\eta)+12\gamma_1^2+(12\chi^2-16\sqrt{3}\chi+20)\gamma_2^2
 \nonumber\\&&+[(16\sqrt{3}-24\chi)\gamma_2+24\eta-12\chi^{2}+16\sqrt{3}\chi-8]\gamma_1
 \nonumber\\&&+[12\chi^3-24\sqrt{3}\chi^2+(48-24\eta)\chi+16\sqrt{3}\eta-8\sqrt{3}]\gamma_2
,\nonumber
 \end{eqnarray}
where $\gamma_1$ and $\gamma_2$ are two real constants. $H_1(\chi,\eta)=6i\chi^3-
 (9i\sqrt{3}+9)\chi^2+(6\sqrt{3}+6i-12i\eta)\chi+(2i\sqrt{3}-6)\eta+2i\sqrt{3}-6$ and $M_1(\chi,\eta)=3\chi^4-8\sqrt{3}\chi^3+(24-12\eta)\chi^2
+(16\sqrt{3}\eta-8\sqrt{3})\chi+4-8\eta+12\eta^2$. When $a=\gamma_1=\gamma_2=1$, the dynamics of solution (\ref{solution}) corresponds to the one in Fig. 1. For $a\neq 1$ and other values of $\gamma_{1,2}$, this solution demonstrates the identical dynamical process except some trivial shifts on the temporal-spatial plane. Performing asymptotic analysis on the exact solution, we find that the two W-shaped solitons will have identical profiles when $\eta \rightarrow \infty$. The hump intensities and valley intensities of them tend to be $4$ and $\frac{5}{8}$ times the CWB intensity value, respectively. This analysis indicates that the two W-shaped localized waves indeed evolve into solitons.
\begin{figure}[t]
\centering
\subfigure[]{\includegraphics[width=85mm,height=65mm]{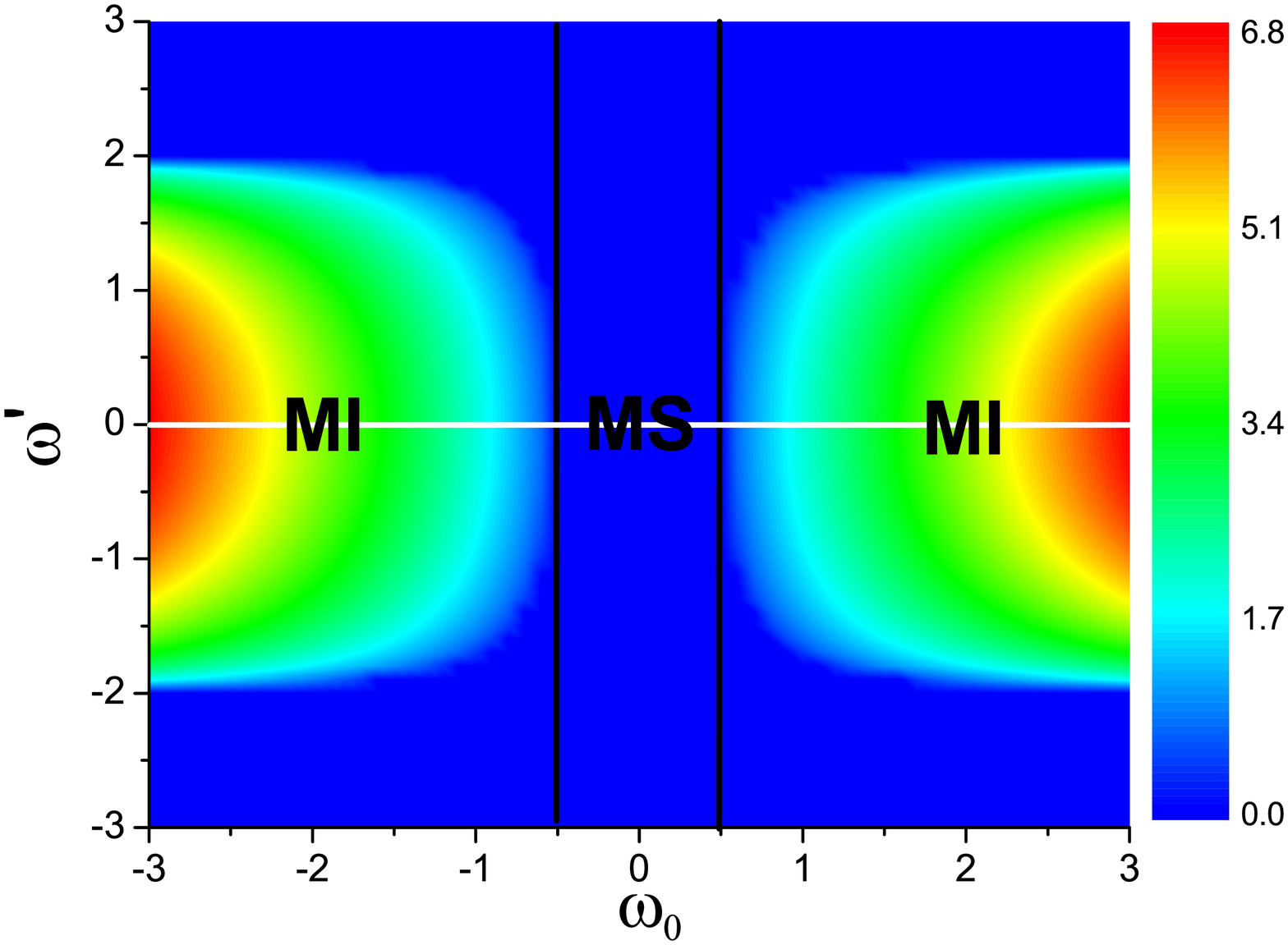}}
\hfil
\subfigure[]{\includegraphics[width=85mm,height=65mm]{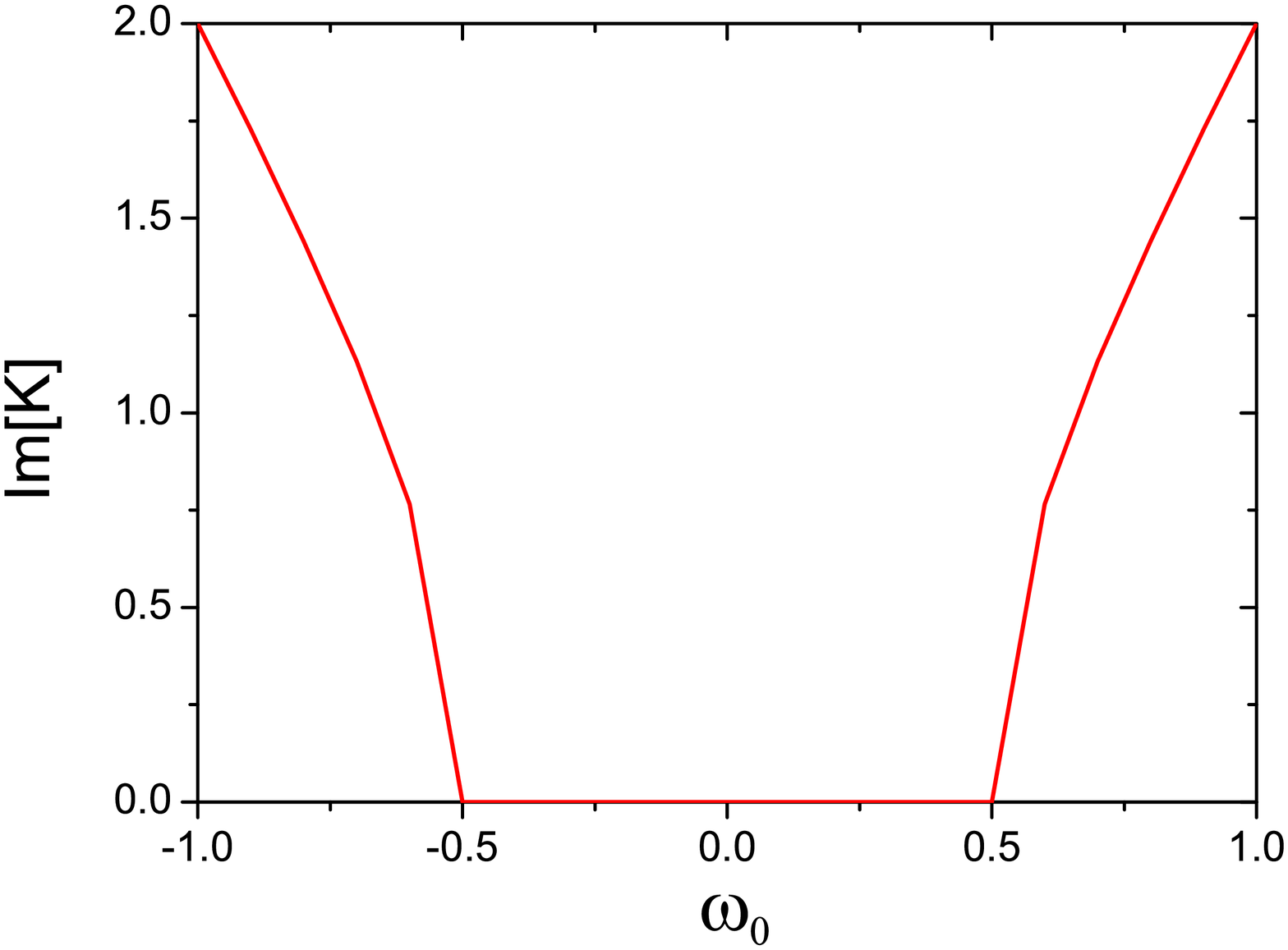}}
\caption{(Color online) (a) The MI gain distribution on background frequency and perturbation frequency plane. The crossing points of black lines and white line are the critical points between MI and MS for resonant perturbation frequency.  (b) The cut plot of MI gain on $\omega'=0$. It is shown clearly that the critical points are at $\pm 1/2$. }\label{fig2}
\end{figure}

As shown in Ref. \cite{Zhao}, the spectrum of W-shaped soliton corresponds to one stable broad spectrum pulse. This means that the two W-shaped solitons correspond to two  broad spectrum pulses, which can be proved by making Fourier transformation on two asymptotic soliton forms of the exact solution. Since there are two W-shaped solitons generated from a small modulation signal, one can expect emerging two stable triangular spectrum and interference fringes. The spectrum evolution of the dynamical process illustrated in Fig. 2 is obtained exactly through the Fourier transformation $F(\omega,\eta)=\frac{1}{\sqrt{2\pi}}\int^{+\infty}_{-\infty}\psi(\chi,\eta)\exp{[-i\omega\chi]}dt$. The background is infinity and it is integrated to be $\delta(\omega-\omega_0)$ (here the background frequency $\omega_0=\frac{1}{2}$), therefore we can eliminate the $\delta$-function by taking the transformation on the varying part.  These properties have been checked (see Fig. 2) and the expected interference pattern on the spectral distribution has been observed.

Now we try to understand the above process for which two W-shaped solitons excited from a small perturbation. To this end, we first perform the linear stability analysis (LSA) on the CWB $\psi_0(\chi,\eta)$ by adding the Fourier modes with small amplitudes \cite{Wright}, i.e., $f_{+}\exp{[i\omega' (\chi-K \eta)]}+f_{-}^\ast\exp{[-i\omega' (\chi-K^\ast\eta)]}$. The MI gain spectrum on background frequency and perturbation frequency space is shown in Fig. 3(a). The results show that there are two different regimes for perturbations on the continuous background, namely,
MI and MS regimes. The critical points for background frequency are  $\omega_0=\pm \frac{1}{2}$, as shown in Fig. 3(b).  For $|\omega_0|\leq\frac{1}{2}$, there exists a MS regime for which the perturbations with arbitrary frequencies are all stable on the CWB. In this regime, stable rational W-shaped soliton solution has been obtained in Ref. \cite{Zhao}. For $|\omega_0|> \frac{1}{2}$, there exists a nonzero MI regime in which the perturbations with frequencies belonging to $[\omega_0-\sqrt{4-\omega_0^{-2}} , \omega_0 +\sqrt{4-\omega_0^{-2}}]$, and the maximum gain of MI will emerge on the ``resonant" line for which the perturbation frequency is equal to the CWB's. The resonant line in MI regime usually corresponds to RW excitation. In this regime, RW solution which possesses double humps has been given in Refs. \cite{Chen,rws}. Notably, the two W-shaped solitons here growing from a small perturbation should involve both two regimes. This process is found on the CWB with the critical frequencies $\omega_0=\pm \frac{1}{2}$ which is located on the boundary between the MS and MI regime. We can qualitatively know that the first stage of the process is in the MI regime for which the small modulation signal grows to one high hump and two deep valleys and then it splits. The second stage of the process should evolve in the MS regime, which makes the split localized waves stable and evolve into two W-shaped solitons. This indicates that MI and MS can coexist on the critical background frequency given by the linear stability analyze. The MI analysis on Eq. (\ref{S-S}) shows that the MS regime for low-frequency perturbations does not exist when neglecting the high-order effects. This implies that these high-order effects play important roles in the dynamical process.

\begin{figure}[t]
\centering{\includegraphics[width=0.9\columnwidth]{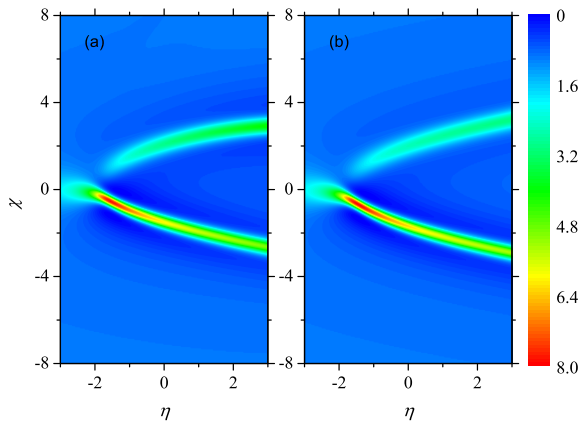}}
\caption{(Color online) (a) The numerical simulation for the dynamics of $|\psi(\chi,\eta)|^2$ evolved from an initial given by the exact solution at $\eta=-3$. (b) The corresponding evolution given by the exact analytical result. It is shown that they agree well with each other, which indicates that the exact signal evolution can survive even with numerical deviations.}
\end{figure}

\begin{figure}[t]
\centering{\includegraphics[width=0.8\columnwidth]{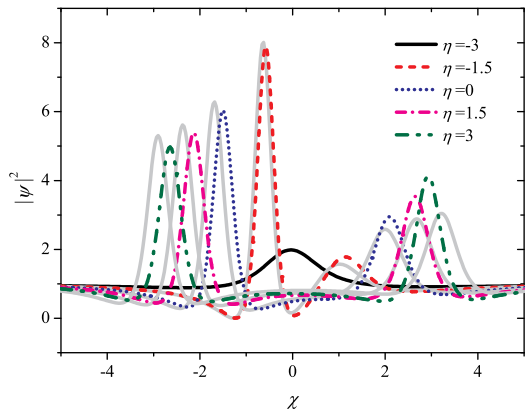}}
\caption{(Color online) The profiles of the  nonlinear waves at different distances. The solid gray curves denote the analytical solutions
while the dashed ones correspond to the numerical results.}
\end{figure}

To see the feasibility of process for W-shaped solitons generation, we numerically evolve Eq. (\ref{S-S-1}) from an exact initial signal with $\eta=-3$ by adopting the split-step Fourier method and the results for $|\psi|^2$ are shown in Fig. 4(a) compared with the analytical solution [see Fig. 4(b)]. Figure 5 demonstrates the profiles comparison. It is shown that the simulation results agree well with the analytical ones even with numerical deviations. A bit differences for the humps' locations and the intensities of the humps is caused by the periodic boundary condition induced reflection effects. The numerical simulation from the initial small signal with some additional noise still evolves to two W-shaped solitons. This results suggest that the exact process might be realized experimentally. It must be mentioned that the W-shaped solitons discussed here are grown from a small modulation signal on the CWB, which are different from the RW and W-shaped solitons reported in earlier studies \cite{Zhao,rws,Chen}. Furthermore, we test the stability of the generated W-shaped solitons with adding a white noise $0.01\ Random(\chi)$ ($\chi\in [-1,1]$).  As an example, we show the deviations between the numerical result with and without the noise in Fig. 6. We can see that the solitons are stable against the noise even with the Raman term.

\begin{figure}[t]
\centering{\includegraphics[width=0.85\columnwidth]{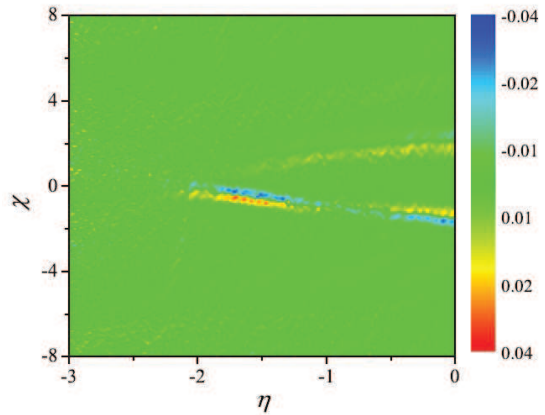}}
\caption{(Color online) The deviations between $|\psi_1|^2$ and $|\psi_2|^2$. The numerical evolutions are from the initial conditions that $\psi_1(\chi,-3)=\psi(\chi,-3)$ and $\psi_2(\chi,-3)=\psi(\chi,-3)+0.01\ Random[\chi]$. It is shown that the deviations are lower than $0.04$. }
\end{figure}

\section{Conclusion and discussion}

In summary, we present a striking dynamical process in which two W-shaped solitons  can be generated from a weak modulation signal on the CWB. The numerical simulations indicate that the dynamical process is robust against small noise or perturbations.
The underlying mechanisms of the process has been discussed qualitatively based on the MI analysis. The process involves MI to MS autonomously. This indicates that MI and MS can coexist on the critical background frequency given by LSA. This is a novel nonlinear dynamical behavior, which could exists in many other nonlinear systems, and even nonlinear coupled systems. Notably, the critical frequency located on the boundary between the MS and MI regimes is obtained by LSA. The analysis predicts there is no MI on the critical frequency, but we demonstrate that both MI and MS play key roles in the dynamical process. This means that LSA could not stand precisely on the critical line. The explicit underlying mechanism for the dynamical process still needs new analysis techniques.

Actually, many recent experiments for nonlinear localized waves including Kuanetsov-Ma breather, Akhmediev breather, and Peregrine RW in nonlinear fibers \cite{Kibler,K-M,Dudley} have provided a good platform to investigate complicated dynamics of nonlinear localized waves on CWB. These experimental works suggested that the exact analytical solutions for the simplified NLSE can well describe the evolution of optical fields even with non-ideal initial conditions. However, the ideal initial conditions is important to study the NLSE with high-order effects and our present work could provide the initial conditions for generating W-shaped solitons in experiment from weak modulation on CWB. This has been checked numerically even for the initial conditions with small noises. In particular, our theoretical results can be used to qualitatively understand the experiment in a nonlinear fiber with TOD and self-steepening effects \cite{Hammani}. In this experiment one Peregrine RW split into two lower-amplitude localized solitons and the pulse splitting process was explained by high-order MI based on a high-order Akhmediev breather solution of the simplified NLSE \cite{Erkintalo}. Here, we provide another perspective for understanding the pulse splitting process and predict the quasi-stability of the splitting pulses with some certain high-order effects.

\acknowledgments
This work is supported by the National Science Foundation of China (Contact No. 11405129, 11305120), and the Fundamental Research Funds for the Central Universities of China.

\end{document}